\documentclass[prl,twocolumn,english,superscriptaddress,floatfix]{revtex4}
\usepackage{graphicx}
\usepackage{amsmath}
\usepackage{amssymb}
\usepackage{color}
\makeatletter

\usepackage{babel}
\usepackage{algorithm}
\usepackage{algorithmic}
\newfloat{mybox}{thp}{lp}
\floatname{mybox}{Algorithm}

\newcommand\figref[1]{Fig.~\ref{#1}}

\newcommand\qEA{q_\mathrm{EA}}
\newcommand\Tg{T^{3\mathrm D}_g}

\makeatother
\begin{document}

\title{Which measures of spin-glass overlaps are informative?}

\author{A. Alan Middleton}
\affiliation{Department of Physics, Syracuse University, Syracuse, New York 13244, USA}

\begin{abstract}
The nature of equilibrium states in disordered materials is often studied
using an overlap function $P(q)$, the probability of two configurations having a similarity $q$.
Exact sampling simulations of a two-dimensional proxy for three-dimensional spin glasses indicate
that common measures of $P(q)$ are inconclusive for systems of linear size $L\le 64$.
Strong corrections result from $P(q)$ being an average over many scales, as
seen in a toy droplet model.
However, the median $\tilde{I}(q)$ of the integrals of  sample-dependent $P(q)$ curves
shows promise for deciding the large size behavior. \end{abstract}
\pacs{}
\maketitle

Materials with quenched disorder have equilibrium and dynamic behaviors that differ
qualitatively from those for homogeneous materials.
The lack of translation invariance leads to a more complex free energy
``landscape'' and to very slow, i.e., glassy, dynamics.
In order to explain the glassiness that is observed in experiment
and predicted by simulations, it is necessary to understand the equilibrium states
approached by relaxation.
The most studied model of disordered systems in statistical mechanics is the spin glass,
where the couplings between nearby spins are randomly set to be ferromagnetic
or antiferromagnetic.
This model serves as a primary example for studying a wide range of disordered materials and
more general problems, including hard optimization problems and neural networks \cite{SGbook1,SGbook2}.
Theoretical characterizations of spin glass equilibrium states
include the replica symmetry breaking (RSB) description \cite{ParisiPRL}, which is based
on the solution of mean-field models \cite{SKmodel}, and the droplet picture \cite{McMillan,BrayMoore,FisherHuse},
which is based on a scaling description.
These each rely on quite distinct pictures of equilibrium: in the former, a countably infinite number of thermodynamic
states are relevant, while in the latter there is a single state, up to global symmetries. Though
RSB-based predictions for free energies have been rigorously verified in mean field models \cite{Talagrand},
analytic verification of either characterization has not been possible in three dimensions \cite{ADNS2010}
and numerical work is generally very difficult due to the glassy dynamics.

To reliably determine the thermodynamic states in models of disordered
materials, it would be useful to extrapolate simulation results
to the large system limit with confidence.
One characterization of the structure of states is the probability
distribution $P(q)$, where $q$ is a measure of the similarity of two equilibrium configurations and
the distribution is an average over disorder realizations $\mathcal J$.
This distribution is not a complete characterization of the equilibrium states \cite{FisherHusePq,NewmanStein1},
but numerical calculation of $P(q)$ has been used as a classic test to discriminate between
many-state pictures such as RSB and pictures such as the droplet model.
For large systems with non-contrived boundary conditions \cite{NewmanStein}, the droplet model predicts that $P(q)$
approaches a single delta function $\delta(q-\qEA)$ for Edwards-Anderson order parameter $\qEA$,
while in many-state pictures such as RSB pictures, the distribution $P(q)$ has additional integrated weight 
over intervals of $q$ that don't include $\qEA$.

This paper presents explorations of which measures of the distribution
$P(q)$ and of the distributions for individual samples, $P_{\mathcal J}(q)$,
can be used to clearly decide between alternate pictures of the thermodynamic states.
The focus is on determining whether there is non-vanishing probability of
seeing macroscopically distinct states in the large system limit.
I present numerical results for the two-dimensional (2D) $\pm J$ Ising spin glass at zero temperature.
This  model has been argued \cite{ThomasHuseMiddleton}
to be a useful proxy for the glassy phase of the three-dimensional Ising (3D) spin glass.
In the 2D $\pm J$ model, a large number of degenerate configurations exist at all scales, but entropic
effects suppress the effects of this degeneracy at large scales, just as 
large scale excitations are suppressed by large free-energy costs within the droplet model
for 3D spin glasses at low enough temperatures $T<T^{3\mathrm D}_g$.
The advantage of the two-dimensional model is that configurations can be exactly sampled \cite{ThomasMiddletonSampling}
for relatively large linear sizes $L\le 256$.
The sample average of $P(q)$ near $q=0$ varies slowly with $L$ for $L\approx 10$,
resembling the results \cite{BanosEtAl,RegerBhattYoung} for three- and
four-dimensional Ising spin glasses that would appear to
support a many-states picture.
However, agreement with droplet scaling expectations is seen in the 2D model for $L>64$.

The slow convergence to the asymptotic behavior 
is argued to result from the contributions from many scales to the distribution $P(q)$.
This is demonstrated using a toy droplet model. In this model, using 
parameters in line with those for the 3D model, $P(q)$ at small $q$ is nearly
constant for $L<512$.
This supports generally being very wary in relying on the averaged $P(q)$ for smaller systems.
A measure of $P_{\mathcal J}(q)$, namely the median $\tilde{I}(q)$
over samples $\mathcal J$ of the cumulative distribution $I_{\mathcal J}(q)=\int_0^q P_{\mathcal J}(q') dq'$,
is proposed to more clearly predict the thermodynamic limit from small numerical samples.

A recently introduced statistic \cite{YKM},
$\Delta(\kappa,q_0,L)$, is also investigated. This statistic is a measure of the
probability of peaks at small $q<q_0$ and is nearly independent of system size in 3D models \cite{YKM}.
In the 2D bimodal simulations, this lack of dependence on $L$ appears to result from a combination of
fewer peaks in $P_{\mathcal J}(q)$ and a sharpening of the peaks as $L$ increases \cite{BilloireEtAl}.
The probability that a peak exceeds a threshold $\kappa$ may then be relatively insensitive to $L$, though
there is only one state. So though
$\Delta$ may in practice distinguish mean field from 3D models \cite{YKM}, this measure requires cautious use.

{\bf Spin glass overlap.} In the Ising spin glass \cite{EAmodel}, the configurations are sets
of spin values $s=\{s_i\}$ indexed by $i=1\ldots N$ and $s_i=\pm 1$ for each spin.
Spin configurations $s$ for a realization ${\mathcal J}$ have a probability
$p(s)\propto e^{-\mathcal{H}_{\mathcal J}(s)/T}$ for
a sample-dependent Hamiltonian $\mathcal{H_J}$.
The spin overlap distribution $P_{\mathcal J}(q)$ for a single sample
${\mathcal J}$ is then just the probability density for the absolute value of the spin overlap
$q=N^{-1}\sum_{i=1}^N s^\gamma_i s^\delta_i $, over independently chosen spin configurations $\gamma$, $\delta$.
As different samples each have distinct couplings $J_{ij}$ between spins $i$ and $j$, $P_{\mathcal J}(q)$ depends on the
coupling realization ${\mathcal J}=\{J_{ij}\}$ as well as the temperature $T$.
The distribution $P(q)$ is the average of $P_{\mathcal J}(q)$ over samples $\mathcal J$.
Using $P(q)$ to characterize states is natural, as $P(q)$ simply gives the probability of
a given overlap $q$ between pairs of equilibrium configurations in a randomly chosen sample. 
Despite the clear difference in the predictions resulting from RSB calculations and droplet models,
it has been difficult to determine numerically the infinite size limit of $P(q)$.
Simulations of spin glasses are difficult due to the extremely
long equilibration times for Markov Chain Monte Carlo methods \cite{NewmanBarkema}, so that
simulations in three dimensions for more than $12^3$ spins are quite challenging.
It has also been argued that simulations just below the spin-glass temperature $\Tg$ suffer from large correlation
lengths that make simulations inconclusive \cite{MooreBokilDrossel}.

{\bf 2D Model.} Distributions $P_{\mathcal J}(q)$ were computed
for a two-dimensional bimodal (i.e., $\pm J$) spin glass in the limit $T\rightarrow 0$.
Model samples have $N=L\times L$ spins arranged on a square lattice with toroidal (periodic) boundary conditions.
A sample realization ${\mathcal J} = \{J_{ij}\}$ is described by couplings $J_{ij}$ which are zero except
for neighboring sites $(i,j)$, where $J_{ij}=\pm 1$ with equal probability. 
The Edwards-Anderson Hamiltonian \cite{EAmodel} is $\mathcal{H}_{\mathcal J}=-\sum_{i=1}^N J_{ij}s_i s_j$.
Pfaffian sampling techniques \cite{ThomasMiddletonSampling} were used to generate configurations $s=\{s_i\}$
with equilibrium probability $Z^{-1}e^{-\beta\mathcal{H}_{\mathcal J}(s)}$, with
$\beta=T^{-1}$ and $Z(\beta)$ the partition function. The value of $\beta$ was set
high enough that equilibrium sampling chooses ground state configurations (Table I lists the values of $\beta$ used
in production runs).
For a given $\mathcal J$, all $m$ (Table I) generated configurations had the same energy $\mathcal{H}_{\mathcal J}$.
Additionally, test runs with $\beta$ at least twice as high as the production $\beta$
gave the same energy for each $\mathcal J$.
This gives high confidence that the production runs generate the ground states for each sample.
For each realization $\mathcal J$, spin overlaps $q$ were computed for all $\frac{m(m-1)}{2}$
pairs of configurations to estimate $P_{\mathcal J}(q)$. To validate
the procedure and to generate configurations,
$1.3\times 10^6\ \mathrm h$ of CPU time were used.

\begin{table}
\caption{Parameters for the two-dimensional $\pm J$ Ising model ground state simulations.  The precision is
the number of bits used for floats, $N_{\mathcal J}$ is the number of disorder realizations, and $m$ is the
number of configurations per realization.}
\begin{ruledtabular}
\begin{tabular}{ccccc}
Size $L$  & $\beta$ ($=1/T$) & Precision & $N_{\mathcal J}$ & $m$\\
8 & 8 & 1024 & $4\times 10^4$ & 1000\\
16 & 8 & 1024 & $4\times 10^4$ & 1000\\
32 & 16 & 1536 & $2\times 10^4$ & 400\\
64 & 16 & 1536 & $1\times10^4$ & 200 \\
128 & 16 & 1536 & $5\times 10^3$ & 200\\
256 & 20 & 2048 & $2\times 10^3$ &50 
\end{tabular}
\end{ruledtabular}
\label{tab:params}
\end{table}

This model was chosen as it has features that are similar to those of the 3D Ising spin glass model for $T<\Tg$.
The entropy difference $\Delta S$ due to changes from periodic to antiperiodic boundaries along
one axis is consistent with the behavior
$\Delta S\sim L^{\theta_S}$, with $\theta_S \approx 0.50$ \cite{JorgEtAl,ThomasHuseMiddleton}.
Zero energy excitations at scale $\ell$ are active
only when their entropy is low enough, which occurs with probability $\sim \ell^{-\theta_S}$, so that spins at large
separation have a preferred relative orientation with finite probability,
allowing for long range spin-glass order \cite{ThomasHuseMiddleton}.
In the droplet model of the 3D Ising spin glass, long range order for $T<\Tg$ results from the free energy cost of domain walls
scaling as $\ell^{\theta}$, for some exponent $\theta>0$ \cite{McMillan,BrayMoore,FisherHuse}.
For the rest of this paper, the notation $\theta_S$ will be replaced with $\theta$ for a uniform
presentation as the entropy dominates over energy;
note that the energy exponent $\theta\approx 0$ for the 
bimodal 2D model \cite{HartmannYoung2001,CHK2004}.
With this replacement, the 2D simulations can parallel the behavior of the three-dimensional spin glass phase, though of course with a distinct value
for $\theta$.

\begin{figure}
\centering
\includegraphics[width=3.4in]{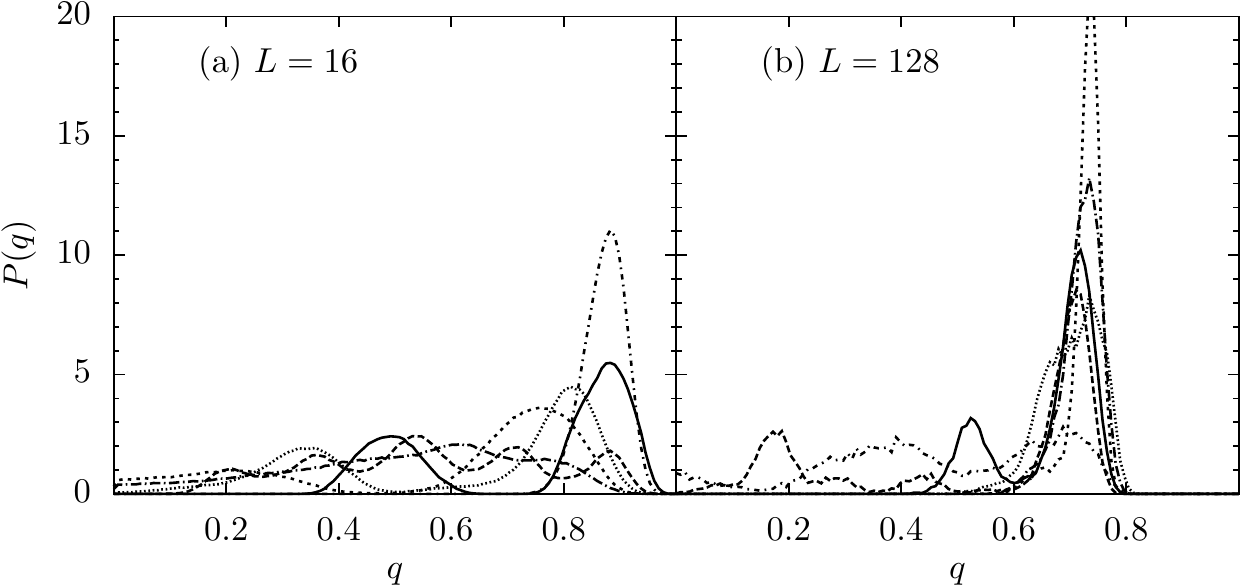}
\caption{Numerically sampled spin overlap distributions $P_{\mathcal J}(q)$ for six realizations ${\mathcal J}$ for the two-dimensional $\pm J$ spin glass
at zero temperature for (a) $L=16$ and (b) $L=128$. The peaks at small $q$ have comparable height in (a) and (b), but become
narrower and so have less total area for larger $L$.
}
\label{fig:Prq}
\end{figure}

\begin{figure}
\centering
\includegraphics[width=3.4in]{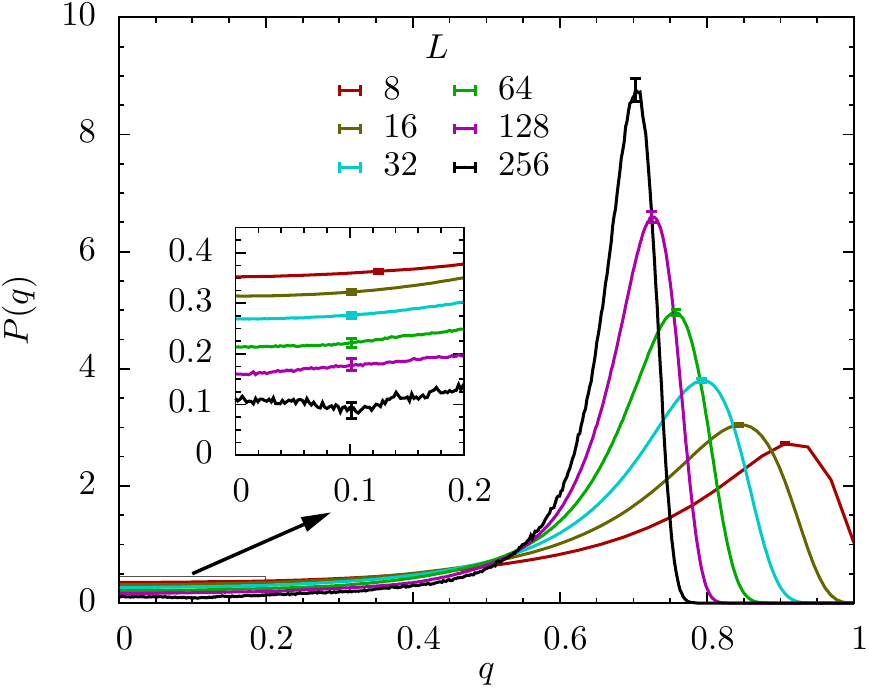} 
\caption{(Color online)
Plots of the sample-averaged $P(q)$ for system sizes $L=8,\ldots,256$.
With increasing $L$, the peaks sharpen and the values at small $q$ decrease, though slowly at smaller $L$.
The error bars show 90\% confidence intervals.
}
\label{fig:avg}
\end{figure}

{\bf Numerical results.} Randomly selected distributions $P_{\mathcal J}(q)$ are shown in \figref{fig:Prq} for the 2D $T=0$
model. For $8\le L \le 256$, most samples
have a peak at $q\approx 0.7-0.8$ and some samples have peaks at smaller $q$.
The peaks become sharper as the system size is increased. The average over realizations $P(q)$ is plotted in \figref{fig:avg}.
The location of the large-$q$ peak can be fit by $q_{\mathrm EA}= 0.645\pm0.01$ with a $L^{-1/\hat{\nu}}$ correction \cite{BanosEtAl}
and a peak height described by $\sim L^{1/\hat{\nu}}$, with $1/\hat{\nu}\approx 0.4(1)$ over one decade in $L$.

\begin{figure}
\centering
\includegraphics[width=3.4in]{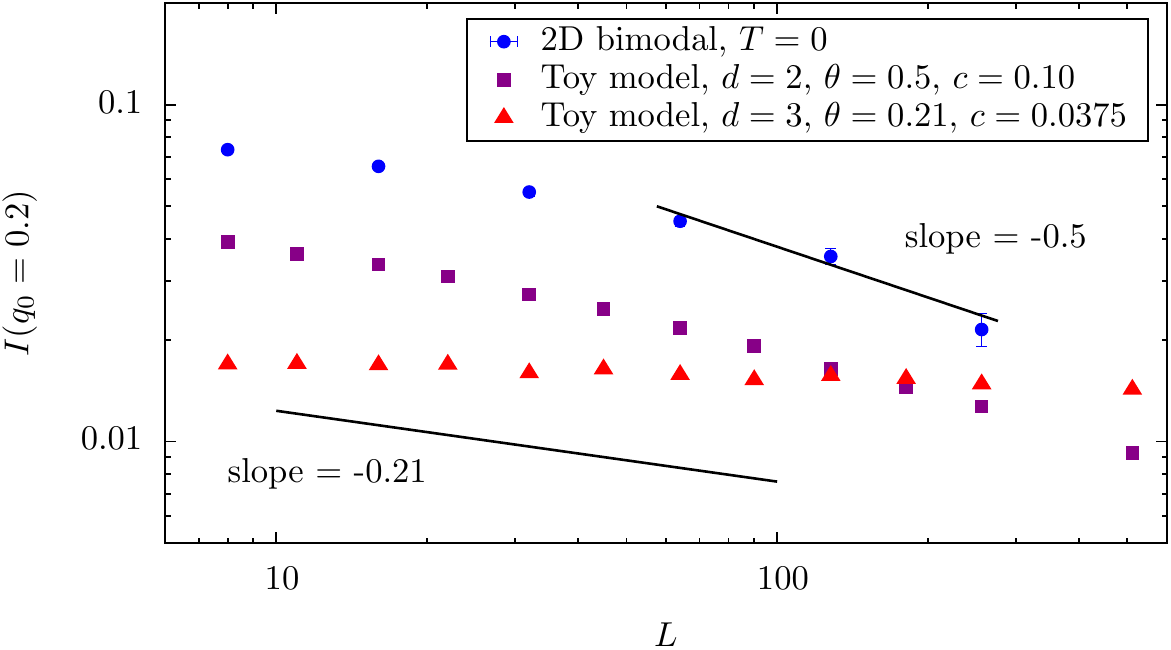} 
\caption{(Color online)
A plot of $I(0.2)$, the cumulative distribution through $q_0=0.2$,   vs.\ $L$, for sampled ground states of the $\pm J$ Ising model
(error bars show 90\% confidence intervals) and toy model simulations.
The expected $I(q_0<\qEA)\sim L^{-\theta}$ holds only for large $L$ in the simulations and the toy model.}
\label{fig:Iscale}
\end{figure}

The computed $P(q)$ for this model varies only slowly with $L$
at small $q$. When comparing $L=8$ with $L=16$, more than $5\times 10^3$ samples
were needed to see a statistically significant difference in $P(0)$.
This slow variation with $L$ resembles the results for the Ising spin glass phase in higher dimensions \cite{BanosEtAl,RegerBhattYoung},
where the lack of visible variation has often been taken as evidence for a many-states picture.
\figref{fig:Iscale} shows the dependence on $L$ of the integrated 
distribution $I(q_0)=\int_0^{q_0}P(q)\,dq$ for $q_0=0.2$. In the droplet picture, droplets of size $\sim L$
appear with frequency $\sim L^{-\theta}$, leading to the expectation
that $I(q_0) \sim L^{-\theta}$ at small $q_0<\qEA$. The data is consistent with this expectation for the range $64 \le L \le 256$,
but the logarithmic slope is less steep for smaller $L$.

{\bf Toy model.} One can study a toy droplet model, such as used in Ref. \cite{HatanoGubernatis}, to explain
these large corrections to scaling (also see \cite{MooreBokilDrossel} for scaling of link overlaps in a hierarchical model).
A given sample $\mathcal J$ is described by a set of active droplets: regions that have fixed relative spin orientations on their interior,
but which are flipped with a probability of exactly $1/2$ with respect to their exterior.
Choosing this flip probability allows nested droplets to be decomposed into independent droplets \cite{HatanoGubernatis}.
The number of independent active droplets with volume between $v$ and $v+dv$
is simply taken to be $n(v)\,dv=cv^{-x}N\,dv$ for some prefactor $c$ and an exponent $x$.
In this calculation, dimensionality $d$
enters only in the relationship between scale $\ell$ and volume $v\sim\ell^d$.
Choosing $x=2+\theta/d$ results in a density of droplets of
scale $\ell$ proportional to $\ell^{-d-\theta}$, in accord with the droplet picture.
To generate a realization $\mathcal J$, the number of droplets of size $v$ is
chosen from a Poisson distribution with mean $n(v)$.
These droplets are then randomly oriented to generate independent configurations.
For $0<\theta<d$, the asymptotic behavior of $P(q)$ near $q=0$ is dominated by the largest droplets and so $P(0)\sim L^{-\theta}$.
However, contributions to $P(0)$ arise from each scale from $L$ down to $1$.
Active droplets at scales below $L$ can be numerous for small $\theta$.
As the number of such scales depends on $L$ and the contribution from several scales below $L$ can strongly contribute
to $P(0)$, the asymptotic behavior $P(0)\sim L^{-\theta}$ may not be evident at small $L$.
Plots of $I(q=0.2)$ for the toy model are included in \figref{fig:Iscale}. The prefactors $c=0.1$ for $(d,\theta)=(2,0.5)$ and
$c=0.0375$ for $(d,\theta)=(3,0.21)$ are chosen to replicate accepted values of $\theta$ and
so that the peaks in the toy model $P(q)$ are near $q=0.8$ for $L\approx 10$.
The magnitude of the local exponent, $|\Delta \ln[I(q)]/\Delta \ln(q)|$, which approaches $\theta$ for
large $L$, exceeds 0.4 only for $L>64$ for the 2D toy model.
For the $(d,\theta)=(3,0.21)$ toy model, the integrated density $I(0.2)$ varies by less than $1\%$ for $8\le L \le 512$,
rather than the factor of $\approx 2.4$ given by asymptotic scaling; the magnitude of the
local exponent exceeds $0.1$ for $L>3\times 10^3$ and it exceeds $0.15$ only for $L>3\times 10^6$.
More detailed models can allow for lattice effects, the effect of boundary conditions
on the density of large droplets, flip probabilities that are not $1/2$,
and interference between overlapping droplets, but such elaborations do not change the
qualitative conclusion of large corrections for small $\theta/d$.

\begin{figure}
\centering
\includegraphics[width=3.4in]{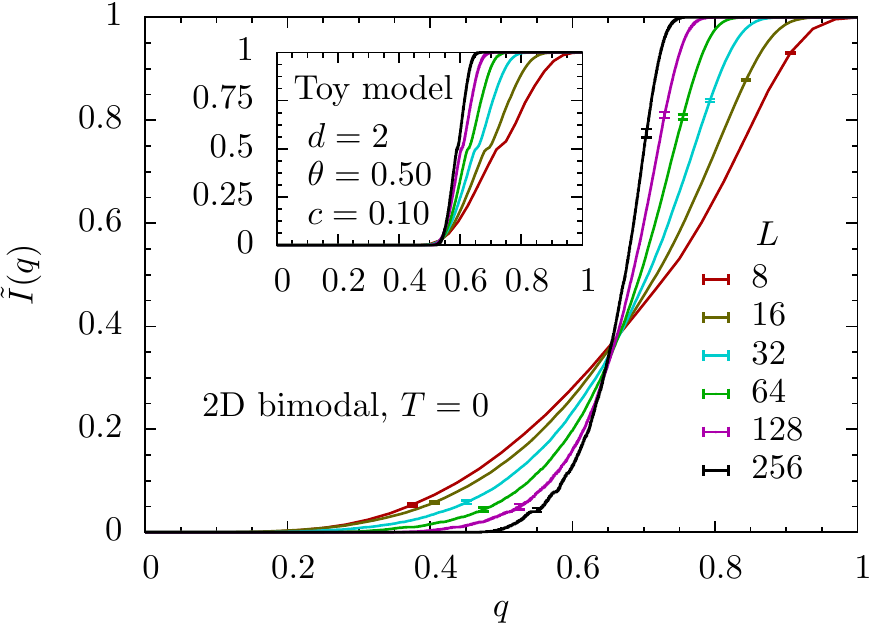}
\caption{(Color online)
Plot of the median $\tilde{I}(q)$ over realizations ${\mathcal J}$ of the cumulative distribution $I_{\mathcal J}(q)$
for ground states of the 2D $\pm J$ Ising model. The values of $L$ range from $L=8$ for the broadest
curve to $L=256$ for the narrowest curve. The error bars show 90\% confidence intervals.  The small values of $\tilde{I}(q)$ at small $q$ do not agree with a many states picture
for this model. Inset: toy model results for the same statistic.
}
\label{fig:median}
\end{figure}

\begin{figure}
\centering
\includegraphics[width=3.4in]{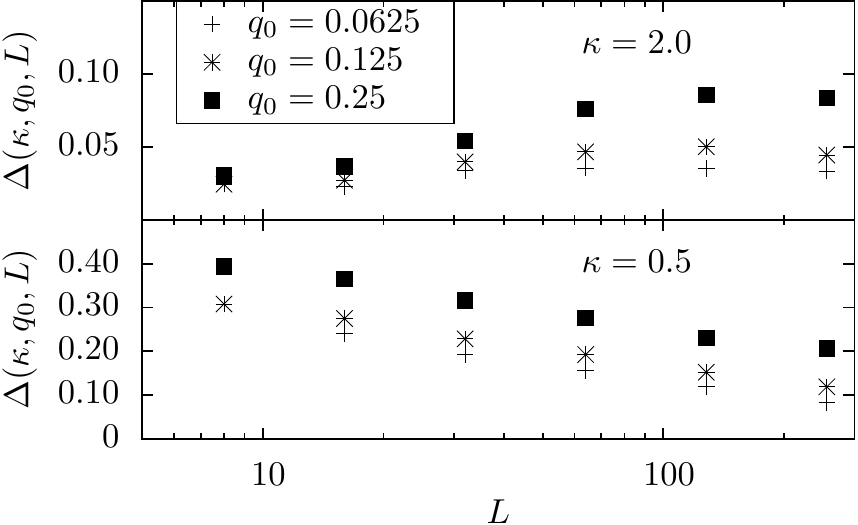}
\caption{
Plots of $\Delta(q_0,\kappa,L)$, the probability that $P(q)$ exceeds $\kappa$ in the range $0\le q \le q_0$,
 vs.\ $L$ for the 2D $\pm J$ Ising spin glass ground states,
with $\kappa=0.5$ and 2 and $q_0=0.0625$, $0.125$, and $0.25$.
For fixed $q_0$ and $\kappa$, this statistic can rise or decline with increasing $L$, due to the competition between the diminishing
weight of peaks and the narrowing of the peaks.
}
\label{fig:Delta}
\end{figure}

{\bf Sample-dependent statistics.}
In order to more easily use spin overlap data to decide the nature of states, a different statistic is now proposed.
This statistic, the sample median $\tilde{I}(q)$ of the integrated distribution function
$I_{\mathcal J}(q)=\int_0^q P_{\mathcal J}(q')dq'$, can be evaluated at each $q$.
If $\tilde{I}(q)$ at small $q$ and large $L$ vanishes, a many states picture cannot hold, as the fraction of samples that
are macroscopically different from each other would necessarily vanish. In the RSB picture,
for example, the majority of samples have a positive $I(q)$ at fixed small $q$ and large enough $L$.
The results for the 2D bimodal
model are shown in \figref{fig:median}: $\tilde{I}(q)$ quickly vanishes with increasing $L$ at small $q<0.2$, though the convergence
is slower at higher $q$. In the toy model with the 2D and 3D parameters used here, $\tilde{I}(q)$ is indistinguishable from zero
for $q<0.4$. The rise of $\tilde{I}(q)$ narrows with increasing $N$, with a
shift to lower $q\rightarrow \qEA$ as $N\rightarrow\infty$.
One could also generalize $\tilde{I}(q)$ to $I_x(q)$, the cumulative value $I_x(q)$ exceeded by a fraction $x$ of the samples, with
$\tilde{I}(q)=I_{1/2}(q)$, to probe the distribution of $I_{\mathcal J}(q)$ in more detail.

A recently introduced \cite{YKM} statistic is $\Delta(q_0,\kappa,L)$, the fraction of samples
where $P_{\mathcal J}(q)$ exceeds $\kappa$ in the range $0\le q \le q_0$.
This statistic was used to search for the multiple peaks in $P(q)$ that are
expected to arise in a many states picture \cite{AspelmeierEtAl2008}.
The value of $\Delta(q_0,\kappa,L)$ was found \cite{YKM} to be roughly constant in $L$ for small $q_0$ and $\kappa$
for the three-dimensional Ising spin glass and to rise with $L$ in the mean field Sherrington-Kirkpatrick model \cite{SKmodel}.
The results of the
simulations for two-dimensional Ising glass ground states are plotted in \figref{fig:Delta}. Behavior very similar to that
in the three-dimensional Ising spin glass is seen: $\Delta$ is nearly constant with $L$ and can rise or fall slowly depending on the
choice of $\kappa$.
Though the peaks in $P_{\mathcal J}(q)$ become less frequent and have less integrated weight as $L$ increases, the peak height does
not vary rapidly with $L$, so that this measure can be
insensitive to the diminishing integrated weight over intervals of $P_{\mathcal J}(q)$.

{\bf Conclusions.}
Numerical studies of the distribution of configuration overlaps in
the two-dimensional bimodal ($\pm J$) Ising model at zero temperature and of a toy droplet model
have been carried out.
The 2D Ising model results show clear evidence for a single pair of thermodynamic states, despite
the complexity resulting from energetic degeneracies at each scale \cite{Hartmann2Ddroplet}.
But both models show very strong finite size corrections to scaling in most overlap statistics. These corrections are strong enough to
make system averages of $P(q\approx 0)$ nearly constant, instead of scaling as $L^{-\theta}$, for $L<500$ 
in the toy model when using parameters consistent with the 3D Ising spin glass.
In the two dimensional $\pm J$ model, system sizes $L>64$  are required to see behavior approaching the scaling predictions.
The corrections seen in the toy droplet model result
from the slow change with scale of the contributions to $P(q)$ and the
sum over scales from $1$ to $L$.
A parallel situation is that of
strong corrections to scaling of average droplet energies that occur when averaging over
droplet sizes \cite{MiddletonFS}; domain walls introduced by boundary
condition changes \cite{McMillan,BrayMoore} show much smaller corrections to scaling.
It may be useful to use similar $L$-scale measures of overlaps to clarify the convergence of $P(q)$ in small systems.
A proposed statistic based instead on individual samples, the median $\tilde{I}(q)$ over samples ${\mathcal J}$ of the cumulative distribution,
is found to be nearly zero at small $q$ even for small samples, allowing such samples to provide
much more convincing evidence for a single pairs picture.
It would be of interest to employ this statistic to compare the mean-field Sherrington-Kirkpatrick model
and the three-dimensional Ising spin glass model.

This work was supported in part by the National Science Foundation grant DMR-1006731. 
This work was carried out primarily using the Syracuse University HTC Campus Grid, a computing
resource of approximately 2000 desktop computers supported by Syracuse University. I thank the Aspen Center
for Physics, supported by NSF grant 1066293, where discussions helped inspire this work, and Creighton
Thomas and Jon Machta for specific discussions.


\begin{thebibliography}{10}

\bibitem{SGbook1} M. M\'ezard and A. Montanari,  \emph{Information, Physics, and Computation},
(Oxford University Press, 2009).

\bibitem{SGbook2} D. L. Stein and C. M. Newman, \emph{Spin Glasses and Complexity} (Princeton University Press, 2012).

\bibitem{ParisiPRL} G. Parisi, Phys. Rev. Lett. {\bf 50}, 1946 (1983).

\bibitem{SKmodel}
D. Sherrington and S. Kirkpatrick, Phys. Rev. Lett. {\bf 35}, 1972 (1975).

\bibitem{McMillan} W. L. McMillan, J Phys. C {\bf 17}, 3179 (1984).
  
\bibitem{BrayMoore} A. J. Bray and M. A. Moore, J. Phys. C {\bf 17}, L463 (1984).

\bibitem{FisherHuse} D. S. Fisher and D. A. Huse, Phys. Rev. Lett. {\bf 56}, 1601 (1986);
Phys. Rev. B {\bf 38}, 386 (1988).

\bibitem{Talagrand} M. Talagrand, Ann. Math {\bf 163}, 221 (2006).

\bibitem{ADNS2010} L.-P. Arguin, M. Damron, C. Newman, D. Stein,
Commun. Math. Phys. {\bf 300}, 641 (2010).

\bibitem{FisherHusePq} D. A. Huse and D. S. Fisher, J. Phys. A {\bf 20}, 997 (1987).

\bibitem{NewmanStein1} 

\bibitem{NewmanStein}C. M. Newman and D. L. Stein, Phys. Rev. Lett. {\bf 76}, 4821 (1996).

\bibitem{ThomasHuseMiddleton}C. K. Thomas, D. H. Huse, and A. A. Middleton,
	Phys. Rev. Lett. \textbf{107}, 047203 (2011).
	
\bibitem{ThomasMiddletonSampling}C. K. Thomas and A. A. Middleton, Phys. Rev. E
  \textbf{80}, 046708 (2009); {\tt http://arxiv.org/abs/1301.1252}.

\bibitem{BanosEtAl} R. A. Ba\~nos, et al., J. Stat. Mech., P06026 (2010).

\bibitem{RegerBhattYoung} J. D. Reger, R. N. Bhatt, and A. P. Young, Phys. Rev. Lett {\bf 64}, 1859 (1990).

\bibitem{YKM} B. Yucesoy, H. G. Katzgraber, and J. Machta,
Phys. Rev. Lett. {\bf 109}, 177204 (2012).

\bibitem{BilloireEtAl} A. Billoire, et al, {\tt http://arxiv.org/abs/1211.0843}.

\bibitem{EAmodel}
S.. Edwards and P. W. Anderson, J. Phys. F {\bf 5}, 965 (1975).

\bibitem{NewmanBarkema} M. E. J. Newman and G. T. Barkema,  \emph{Monte Carlo Methods in Statistical Physics}
(Oxford University Press, 1999).

\bibitem{MooreBokilDrossel} M. A. Moore, H. Bokil, and B. Drossel, Phys. Rev. Lett. {\bf 81}, 4252 (1998).
  
\bibitem{JorgEtAl}T. J\"{o}rg, J. Lukic, E. Marinari, and O. C. Martin,
  Phys. Rev. Lett. \textbf{96}, 237205 (2006).
  
  \bibitem{HartmannYoung2001} A. K. Hartmann and A. P. Young, Phys. Rev. B {\bf 64}, 180404(R) (2001).
  
  \bibitem{CHK2004} I. A. Campbell, A. K. Hartmann, H. G. Katzgraber, Phys. Rev. B {\bf 70}, 054429 (2004).
  
\bibitem{HatanoGubernatis} N. Hatano and J. E. Gubernatis, Phys. Rev. B \textbf{66}, 054437 (2002).

\bibitem{AspelmeierEtAl2008} T. Aspelmeier, A. Billoire, E. Marinari, and M. A. Moore, J.
Phys. A: Math. Theor. {\bf 41}, 324008 (2008).

\bibitem{Hartmann2Ddroplet} A. K. Hartmann, Phys. Rev. B {\bf 77}, 144418 (2008).
  
\bibitem{MiddletonFS} A. A. Middleton, Phys. Rev. Lett. \textbf{83}, 1672 (1999).

\end{thebibliography}
\end{document}